\documentclass[journal,draftcls,onecolumn]{IEEEtran}

\author{Daoud Clarke}
\date{\today}
\title{Riesz Logic}

\usepackage{amsmath}
\usepackage{stmaryrd}
\usepackage{cmll}
\usepackage{graphicx}
\usepackage{caption}
\usepackage{subcaption}
\usepackage{pstricks}
\usepackage{fancyvrb}
\usepackage{amsthm}
\usepackage{verbatimbox}
\usepackage{cite}

\theoremstyle{definition}
\newtheorem{definition}{Definition}

\newcommand{\interp}[1]{\llbracket #1 \rrbracket}

\begin{document}

%% \begin{IEEEkeywords}

%% \end{IEEEkeywords}

\maketitle

\begin{abstract}
We introduce Riesz Logic, whose models are abelian lattice ordered
groups, which generalise Riesz spaces (vector lattices), and show
soundness and completeness. Our motivation is to provide a logic for
distributional semantics of natural language, where words are
typically represented as elements of a vector space whose dimensions
correspond to contexts in which words may occur. This basis provides a
lattice ordering on the space, and this ordering may be interpreted as
``distributional entailment''. Several axioms of Riesz Logic are
familiar from Basic Fuzzy Logic, and we show how the models of these
two logics may be related; Riesz Logic may thus be considered a new
fuzzy logic. In addition to applications in natural language
processing, there is potential for applying the theory to neuro-fuzzy
systems.\footnote{Submitted to IEEE Transactions on Fuzzy Logic,
  Copyright 2014 IEEE}
\end{abstract}


\IEEEpubid{0000--0000/00\$00.00~\copyright˜2014 IEEE}

\section{Introduction}

\IEEEPARstart{M}{uch} of the original motivation for fuzzy logic revolved around
linguistic intuitions, for example the notion that ``tall'' is not a
black and white concept, but that there are degrees of
tallness. Indeed, one of the proposed applications for these ideas was
in linguistics \cite{Zadeh:73}. However, these ideas were never
directly adopted by the linguistics or computational linguistics
community. Instead, fuzziness has crept into natural language
semantics research by the widespread adoption of ``distributional
semantics'', in which the meaning of words is determined by the
contexts in which they occur. These techniques typically represent
word meanings as vectors over these contexts, which capture fuzzy
relationships between word meanings.

One question that is now being studied is how these vector based
representations can be related to older, logical representations of
meaning, in which a sentence would typically be translated into a
logical form. The goal of this paper is to show that the vector spaces
used in distributional semantics, Riesz Spaces, can be considered as
models for a logic, which we call Riesz Logic (RL). Our hope is that
this will lead to a confluence of distributional and logical
semantics, opening up new areas of research and new methods of
tackling problems in natural language processing.

%%  Initially,
%% researchers were mainly interested in using these vectors to derive
%% symmetric measures of similarity between word meanings; more recently,
%% it has been proposed that we consider measures that can be interpreted
%% as degrees of entailment.

%% This ``distributional entailment'' makes use of the implicit lattice
%% ordering on the vector space whose meet and join correspond to
%% component-wise minimum and maximum respectively, where the basis is
%% defined by the set of contexts.

%%  Whilst these
%% vectors have been used to give measures of similarity of word
%% meanings, and degrees of entailment, until now they haven't been

%% We introduce Riesz Logic (RL), a variant of fuzzy logic whose models
%% are abelian lattice ordered groups, of which the most familiar
%% examples are vector lattices, or Riesz spaces.

RL has the following inference rules:\\
\begin{minipage}{0.49\columnwidth}
\begin{gather}
  \tag{MP} \frac{\phi, \phi \rightarrow \psi}{\psi}
\end{gather}
\end{minipage}
\begin{minipage}{0.49\columnwidth}
\begin{gather}
  \tag{RI} \frac{\phi \rightarrow \psi}{\phi \lor \chi \rightarrow \psi \lor \chi}
\end{gather}
\end{minipage}
\vspace{0.3cm}\\
and axioms:
\begin{align}
  \tag{R1a} &(\phi \rightarrow \psi) \rightarrow ((\psi \rightarrow \chi)
  \rightarrow (\phi \rightarrow \chi))\\
  \tag{R1b} &((\psi \rightarrow \chi) \rightarrow (\phi \rightarrow
  \chi)) \rightarrow (\phi \rightarrow \psi)\\
  \tag{R2} &\phi \rightarrow \phi \lor \psi\\
  \tag{R3} &\phi \lor \psi \rightarrow \psi \lor \phi,\\
  \tag{R4} &(\phi \lor \psi)\lor \psi \rightarrow \phi \lor \psi\\
  \tag{R5a} &0 \rightarrow (\phi \rightarrow \phi)\\
  \tag{R5b} &(\phi \rightarrow\phi) \rightarrow 0\\
  \tag{R6a} &((\phi \rightarrow \psi)\lor 0 \rightarrow (\psi \rightarrow \phi) \lor 0) \rightarrow (\psi \rightarrow \phi)\\
  \tag{R6b} & (\psi \rightarrow \phi) \rightarrow ((\phi \rightarrow \psi)\lor 0 \rightarrow (\psi \rightarrow \phi) \lor 0)
\end{align}
In this paper we prove the soundness and completeness of this logic
with respect to abelian lattice ordered groups, which generalise Riesz
spaces, with formulas interpreted as asserting positivity. In doing
this, we relate RL to the Logic of Equilibrium, known as BAL
\cite{Galli:04}.

%  In particular there is no ``True'' or
% ``False'' constant, instead there is a zero value which indicates
% maximal uncertainty about a proposition. We demonstrate that our logic
% is equivalent to the logic BAL of

\section{Background}

\subsection{Distributional Semantics}

Distributional semantics (see \cite{Turney:10} for a comprehensive
overview) is founded on the idea that the meaning of words can be
determined by observing the contexts in which they occur. This idea
has its origin in the work of Firth \cite{Firth:57} and Harris
\cite{Harris:68}, and the philosophy of Wittgenstein
\cite{Wittgenstein:53}. This idea has lead to techniques which analyse
large text corpora to build word representations. For example, Figure
\ref{fig:fruit} shows a sample of occurrences of the word ``fruit'' in
the British National Corpus. Word representations built from such
corpora are typically vectors describing the frequency with which the
word occurs in different contexts. Depending on the application, the
set of contexts which form the basis for the vector space will vary:
\begin{itemize}
\item Document identifiers: the context of a word is treated as the ID
  of the document in which it occurs.
\item Other words: a word is considered to cooccur with another word
  if they are seen together within a window of a fixed number of
  words, or within the same sentence.
\item Grammatical relations: sentences may be parsed to give
  dependency relations between words, and these relations treated as
  contexts.
\end{itemize}
Table \ref{table:fruit} shows hypothetical occurrences of a few terms
where document identifiers have been used as the contexts.

These raw frequency vectors are then typically processed in a variety
of ways to reliably determine relationships between words. A typical
system will employ one or more of the following techniques:
\begin{itemize}
\item Stopword removal or feature selection to identify contexts that
  provide useful contributions to the word's meaning
\item Replacing frequency counts with derived statistics such as
  pointwise mutual information or TF-IDF (term frequency-inverse
  document frequency)
\item Dimensionality reduction such as random projection or truncated
  singular value decomposition.
\end{itemize}
After these processing stages, instead of being integer frequency
counts, word vectors are now typically real-valued, and may even
contain negative values, for example if pointwise mutual information
is used (although it is also common to set the negative components to
zero). There is also typically still a preferred basis for the vector
space, although the meaning of the basis vectors may no longer be tied
to individual contexts, for example if dimensionality reduction has
been performed. Instead, these dimensions are often considered to
correspond to hidden ``latent'' aspects of meaning.

\begin{figure*}
\begin{center}
\begin{BVerbatim}[fontsize=\scriptsize]
end some medicine for her, but she will need fruit and  milk, and some other special things that
our own. Here we give you ideas for foliage, fruit and  various festive trimmings that you can i
part II). However, other strategies can bear fruit  and are described under three sections which
       supper  tomatoes, potato chips, dried fruit and cake. And  they drank water out of tea-cu
erent days, as  the East Berliners queue for fruit and cheap stereos, a Turkish  beggar sleeps i
dening; and  Pests -- how to control them on fruit and vegetables. Both are  produced by the Hen
me,"Silver Queen" is male so will never bear fruit    At the opposite end of the prickliness sca
 lifted away    Like an orange lifted from a fruit-bowl    And darkness, blacker    Than an oil-
ed in your  wreath. Christmas ribbon and wax fruit can be added for colour.  Essentials are scis
e you need to start developing your very own fruit  collection    KEEPING OUT THE COLD    Need e
ly with Jeyes fluid    THE KITCHEN GARDEN    FRUIT    Cut out cankers on fruit trees, except tho
wn and watered    AUTUMN HUES    Foliage and fruit enrich the autumn garden, whether glowing  th
- have forgotten  the maxim: " tel arbre tel fruit ". If I were  willing  to  unstitch the past 
 of three children of Alfred Roger Ackerley, fruit importer  of London, and his mistress, Janett
rful didactic spirit, much that was to  bear fruit in his years as a mature artist. Although thi
e all made with natural vegetable, plant and fruit ingredients  such as chamomile, kukai nut and
ack in the soup.    He re-visits the Copella fruit juice farm in Suffolk, the  business he told 
rategic relationship" with Lotus, the first  fruit of which is a mail gateway between Office and
, choose your plants  carefully to enjoy the fruit of your labour all year round.    PLACES TO V
 and I love chips.  Otherwise I'll nibble on fruit or something to convince myself  that I'm eat
 tone and felt the  softness and warmth of a fruit ripening against a wall? If she  had she migh
ol place to set. Calories per  slice: 395    Fruit Scones with cinnamon Butter    (makes 12)    
ought me water. Another monster gave me some fruit  to eat. A few monsters lay against my body a
ney fungus.    Cut out diseased wood on most fruit trees    VEGETABLES    Continue winter diggin
age and chafing.    Remove old, unproductive fruit trees by cutting them down to  shoulder heigh
ITCHEN GARDEN    FRUIT    Cut out cankers on fruit trees, except those on peaches, plums  and ch
ps  remain, then stir in the sugar and dried fruit. Using a round-  ended knife, stir in the mil
 of a homeland, well others dream too,    De fruit was forbidden an now yu can't chew,    How ca
onnoisseurs. We take a bite from an unusual  fruit. We come away neither nourished nor ravished,
\end{BVerbatim}
\end{center}
\caption{Occurrences and some context of occurrences of the word \emph{fruit} in the British National Corpus.}
\label{fig:fruit}
\end{figure*}

\begin{table}
\newcommand\T{\rule[-1.2ex]{0pt}{3.7ex}}
\begin{center}
\begin{tabular}{|l|cccccccc|}
\hline
	 	& $d_1$\T	& $d_2$	& $d_3$	& $d_4$	& $d_5$ & $d_6$ & $d_7$ & $d_8$\\
\hline
banana\T	& 2		& --		& --		& --		& 5		& --		& 5		& --\\
apple\T	& 4		& 3		& 4		& 6		& 3		& --		& --		& --\\
orange\T	& --		& 2		& 1		& --		& --		& 7		& --		& 3\\
fruit\T	& --		& 1		& 3		& --		& 4		& 3		& 5		& 3\\
tree\T	& --		& --		& 5		& --		& --		& 5		& --		& --\\
computer\T& --		& --		& --		& 6		& --		& --		& --		& --\\
\hline
\end{tabular}
\caption{A table of hypothetical occurrences of words in a set of documents, $d_1$ to $d_8$.}
\label{table:fruit}
\end{center}
\end{table}

\subsection{Distributional Generality}

Distributional word vectors are often used in applications where it is
enough to know how similar two words are in meaning. For this purpose,
measures of distributional \emph{similarity} are sufficient, for
example, the cosine of the angle between two vectors is one measure
that is often used. However, for applications such as information
retrieval or question answering it is important to know whether it is
likely that one word \emph{entails} another. Recent research has
investigated to what degree it is possible to determine this from word
vectors
\cite{Weeds2004,Geffet2005,Szpektor2008,Kotlerman2010,Weeds:14}. One
proposal supported by these experiments, known as distributional
\emph{generality} or distributional \emph{inclusion}, is the idea that
words with a more general meaning will occur in a wider range of
contexts.

This idea was formalised in \cite{clarke:12} in terms of the lattice
ordering that is implicit in the vector space. Since the vector spaces
used in these applications almost always have a preferred basis, it is
possible to define a lattice ordering, where the meet and join
operations are the component-wise minimum and maximum
respectively. This makes the space a vector lattice, or Riesz space:

\begin{definition}[Partially ordered vector space]\index{vector space!partially ordered|textbf}
A partially ordered vector space $V$ is a real vector space together with a partial ordering $\le$ such that:
\vspace{0.1cm}\\
\indent if $u \le v$ then $u + w \le v + w$\\
\indent if $u \le v$ then $\alpha u \le \alpha v$
\vspace{0.1cm}\\
for all $u,v,w \in V$, and for all $\alpha \ge 0$. Such a partial ordering is called a \textbf{vector space order} on $V$. An element $u$ of $V$ satisfying $u \ge 0$ is called a \textbf{positive element}; the set of all positive elements of $V$ is denoted $V^+$. If $\le$ defines a lattice on $V$ then the space is called a \textbf{vector lattice} or \textbf{Riesz space}.
\end{definition}

\begin{figure*}
\begin{center}
\includegraphics{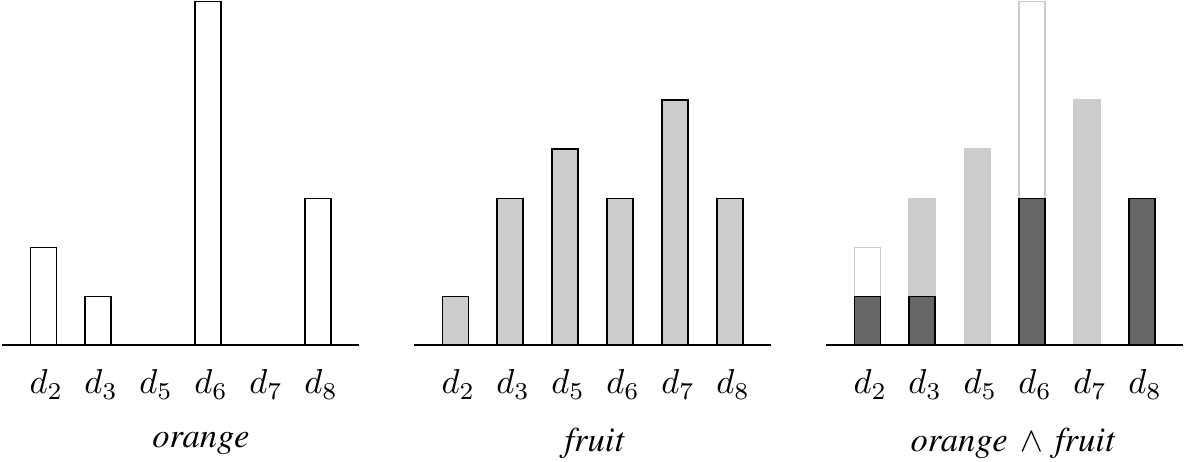}
\caption{Vector representations of the terms \emph{orange} and
  \emph{fruit} and their vector lattice meet (the darker shaded
  area).}
\label{fig:orangefruit}
\end{center}
\end{figure*}

The intuition behind the ordering $\le$ is that it describes a
\emph{distributional entailment}: assuming that $\hat{x}$ and
$\hat{y}$ are distributional vectors of word frequencies for words $x$
and $y$ respectively, then $\hat{x} \le \hat{y}$ means that $y$ occurs
at least as frequently as $x$ in all contexts. Figure
\ref{fig:orangefruit} gives an example of the meet operation for
hypothetical word frequency vectors.

\section{Interpretations}

Later we will prove the soundness and completeness of RL with respect
to abelian lattice ordered groups, which generalise Riesz spaces, with
vector addition and negation forming the group operations.

\begin{definition}[Lattice Ordered Group]
  A partially ordered group is a tuple $\langle G, +, \le\rangle$ such
  that $\langle G, +\rangle$ is a group, and $\le$ is a partial order
  on $G$ such that if $u \le v$ then $u + w \le v + w$ and $w + u \le
  w + v$. If $\le$ is a lattice order, then $G$ is called a lattice
  ordered group. Where there is no confusion, we refer to the lattice
  ordered group $\langle G, +, \le\rangle$ as simply $G$. We denote
  the lattice meet and join by $\land$ and $\lor$ respectively.

  The \textbf{positive part} of $u\in G$ is written $u^+$ and is defined
  as $u^+ = u\lor 0$; its \textbf{negative part} is defined as $u^- =
  (-u)\lor 0$.
\end{definition}

Riesz spaces are abelian lattice ordered groups where the group
operation is vector space addition, and the vector space zero is the
unit of the group.

An interpretation $\langle G, F\rangle$ for RL is an abelian lattice
ordered group $G$ and a function $F$ that maps variables in RL to
elements of $G$. A formula $x$ has the interpretation $\interp{x}$
defined recursively as follows:
\begin{itemize}
\item $\interp{\phi} = F(\phi)$
\item $\interp{x \rightarrow y} = \interp{y} - \interp{x}$
\item $\interp{x \lor y} = \interp{x} \lor \interp{y}$
\item $\interp{0} = 0$
\end{itemize}
Note that the symbols $\lor, \land$ and $0$ are used both as symbols
in the logic (on the left hand side) and in their vector space sense
(on the right hand side).

The formula $x$ is interpreted as asserting that $0 \le
\interp{x}$. Thus, for example, the formula $\phi \rightarrow \psi$ is
interpreted as the assertion $0 \le F(\psi) - F(\phi)$, or $F(\phi)
\le F(\psi)$. A formula $x$ is satisfiable if there is some
interpretation such that $0 \le \interp{x}$; it is a theorem or
tautology if $0 \le \interp{x}$ for all interpretations.

\section{Relation to Fuzzy Logic and Neural Networks}

\subsection{Fuzzy Logic}

Our goal is to show that RL may be viewed as a type of fuzzy logic,
although a non-standard one, both to aid in gaining an intuition for
the nature of the logic, and to demonstrate its potential as a
reasoning system. Firstly, it is worth noting that there is some
overlap between the axioms of RL and Basic Fuzzy Logic (BL): R1a is an
axiom of BL, and R2--R4 hold in BL since $\lor$ is a lattice join;
other axioms are specific to RL.

Most fuzzy logics are interpreted in terms of the real interval
$[0,1]$. Consider the vector lattice of the real numbers (the
single dimensional vector space). This can be mapped to the open
interval $(0,1)$, for example using the logistic function:
$$f(x) = \frac{1}{1 + e^{-x}}$$
The operations $\land$ and $\lor$ (maximum and minimum) behave the
same when their behaviour is translated to this space.

\begin{figure}
  \begin{subfigure}{0.49\textwidth}
    \includegraphics[width=\textwidth]{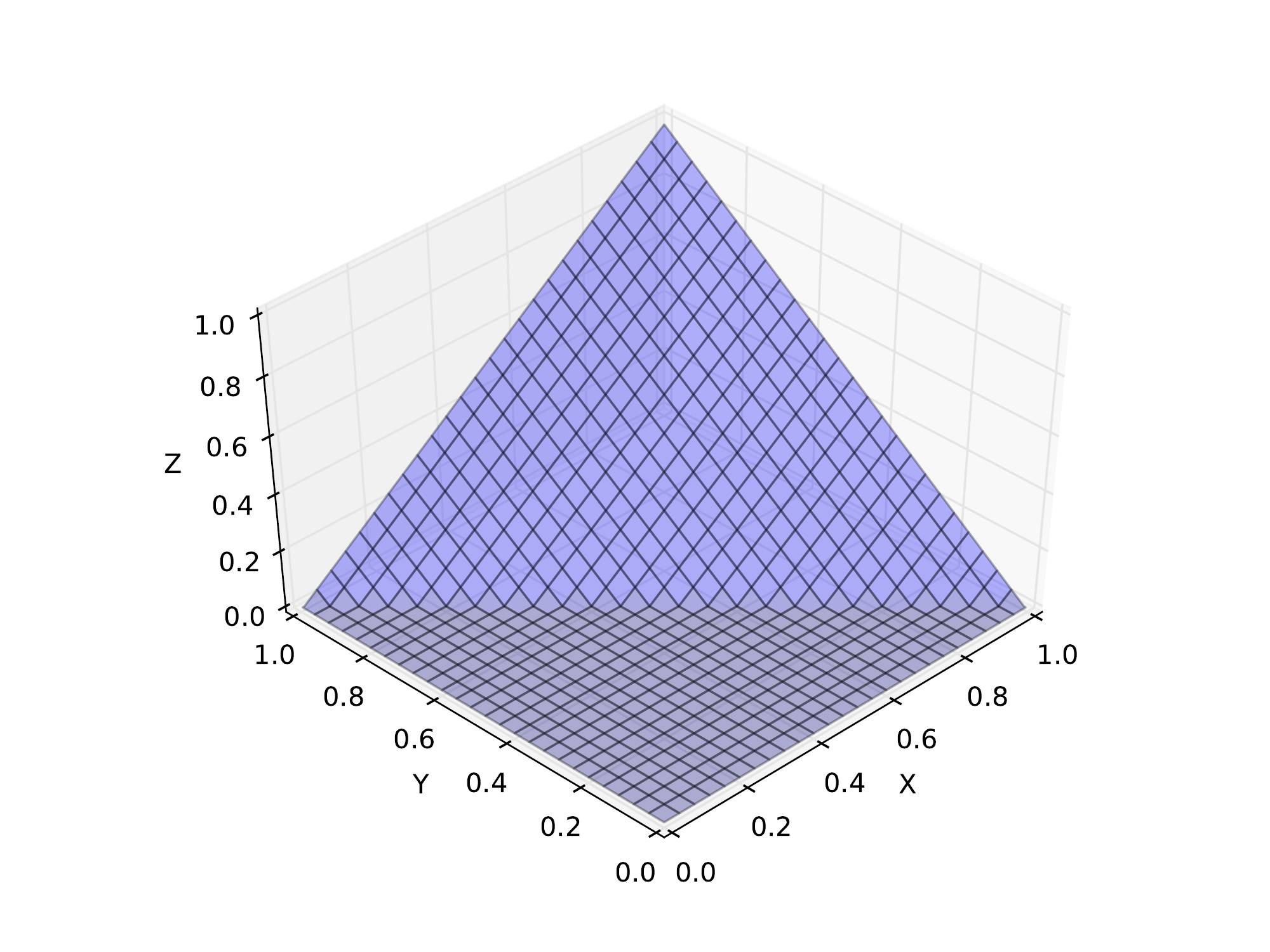}
    \caption{\L ukasiewicz t-norm}
    \label{fig:lukasiewicz}
  \end{subfigure}
  \begin{subfigure}{0.49\textwidth}
    \includegraphics[width=\textwidth]{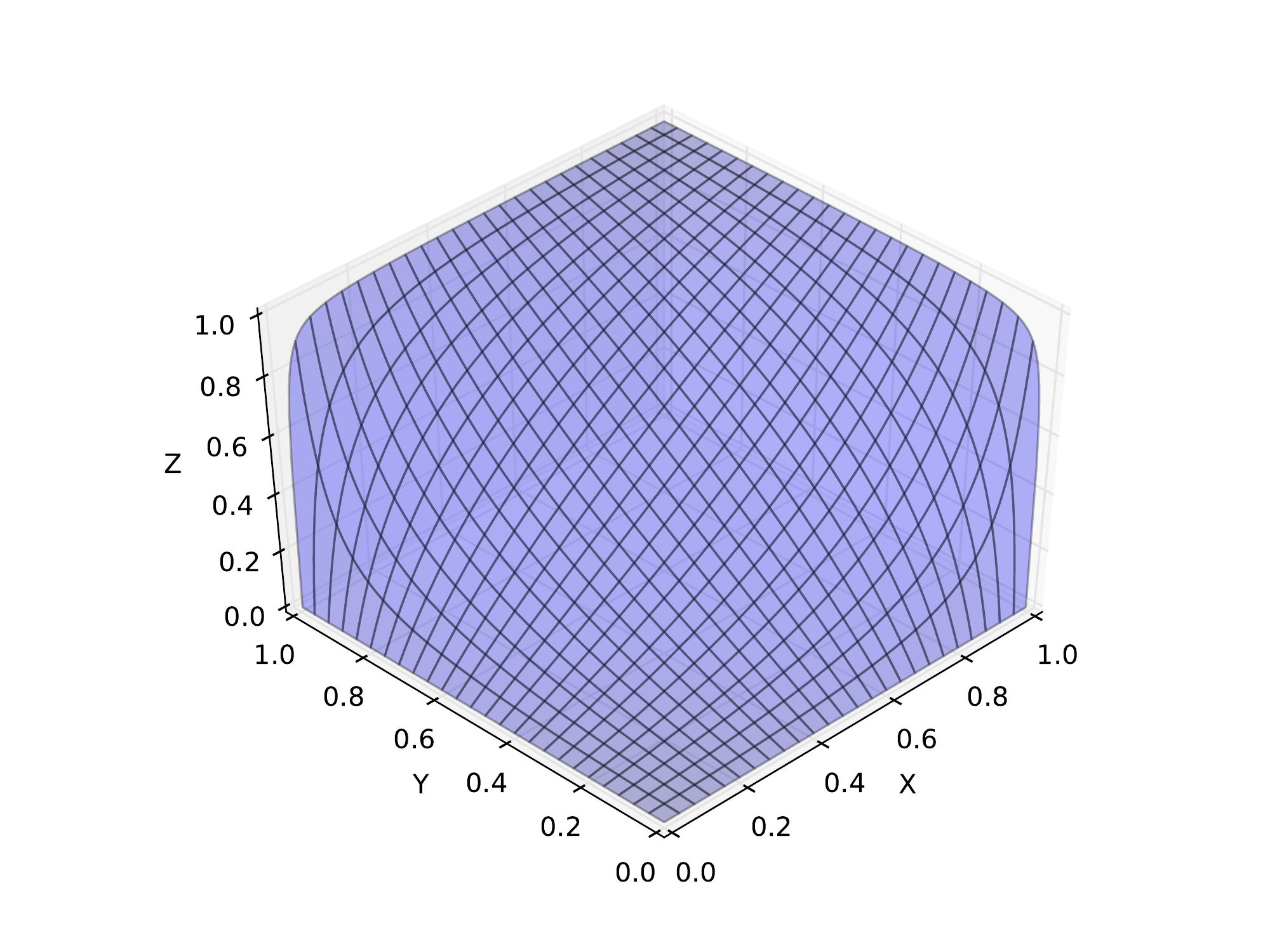}
    \caption{Logistic addition}
    \label{fig:logistic}
  \end{subfigure}
  \caption{Mapping real numbers to the interval $(0,1)$ gives a new
    operation corresponding to addition (\subref{fig:logistic}), which
    bears some similarities to t-norms (\subref{fig:lukasiewicz}).}
  \label{fig:tnorm}
\end{figure}

Many fuzzy logics are derived from T-norms, which have the following
properties:
\begin{align}
\tag{Commutativity} T(a, b) &= T(b, a) \\
\tag{Monotonicity} T(a, b) &\le T(c, d) \text{ if } a \le c  \text{ and } b \le d\\
\tag{Associativity} T(a, T(b, c)) &= T(T(a, b), c)\\
\tag{Identity} T(a, 1) &= a
\end{align}
For example, the \L ukasiewicz T-norm is defined as $T_L(a,b) =
\max\{0,a + b - 1\}$.

A natural question to ask is whether we can define a T-norm for
RL. Vector space addition seems like a natural candidate for this,
because of its similarity to $T_L$. Note that in RL, addition $\oplus$
can be defined by
$$\phi \oplus \psi := (\phi \rightarrow 0) \rightarrow \psi.$$
Addition of two real numbers translates to the interval $(0,1)$ as:
$$T_R(a, b) = \frac{ab}{ab + (1-a)(1-b)}$$
See figure \ref{fig:tnorm} for a three dimensional depiction of $T_L$
and $T_R$. The first three of these properties are satisfied by $T_R$
since they are properties of addition of real numbers, only the
identity property is unsatisfied. Instead $T_R$ is in general
undefined for $a$ or $b$ equal to 1, although clearly for $a \neq 0$,
$T_R(a,1) = 1$, a quite different property from that of T-norms.

Thus RL has some very similar properties to fuzzy logics. The lattice
operations of RL correspond to the weak conjunction and disjunction of
fuzzy logics, whilst vector addition (defined implicitly via the
$\rightarrow$ operation) corresponds to strong conjunction. The major
difference between RL and fuzzy logics is that there are no constants
for ``true'' or ``false'', only the constant 0 representing complete
uncertainty.

\subsection{Neural Networks}

Recent work in deep neural networks \cite{Dahl:13} has shown that
replacing the sigmoid function for activation with a ``rectified
linear unit'' can improve accuracy and training time. A rectified
linear unit is simply the function $f(x) = \max(x, 0)$, which is
equivalent to $f(x) = x^+$ in vector lattice notation. This opens up
the possibility of using RL in combination with neural networks, a
potential new approach to neuro-fuzzy modelling \cite{Lin:91,Jang:95,Abe:12}.

\section{Soundness}

Proving soundness of the logic amounts to proving the validity of the
rule and axioms.

%\subsection{Modus Ponens}

\begin{proof}[MP]
If $0 \le F(\phi)$ and $0 \le F(\psi) - F(\phi)$ then
$0 \le F(\psi)$ by transitivity of $\le$.
\end{proof}

\begin{proof}[RI]
  This follows from simple lattice-theoretic properties: The assertion
  $\phi \rightarrow \psi$ translates to $F(\phi) \le F(\psi)$. As a
  shorthand, let us write $u = F(\phi)$, $v = F(\psi)$ and $w =
  F(\chi)$; we wish to show that $u\le v$ implies $u\lor w \le v \lor
  w$. To see this:
  \begin{align*}
    u &\le v\\
    u\lor v &= v\\
    u\lor v\lor w &= v\lor w\\
    (u\lor w)\lor (v\lor w) &= v\lor w\\
    u\lor w &\le v\lor w.
  \end{align*}
\end{proof}

\begin{proof}[R1]
  Since R1a is the converse of R1b, they may be taken together as
  asserting equality, by the antisymmetry of $\le$. Thus we need to
  show:
  \begin{align*}
    \interp{\phi \rightarrow \psi} & = \interp{(\psi \rightarrow
      \chi) \rightarrow (\phi \rightarrow \chi)}\\
    F(\psi) - F(\phi) & = \interp{\phi \rightarrow \chi} -
    \interp{\psi \rightarrow \chi}\\
    & = F(\chi) - F(\phi) - F(\chi) + F(\psi)\\
    & = F(\psi) - F(\phi).
  \end{align*}
\end{proof}
R2--4 are trivially seen to be properties of the partial ordering. R5
defines the symbol 0 such that $\interp{0}$ is the identity of the
group, which we also denote 0.

\begin{proof}[R6]
\begin{align*}
  \interp{(\phi \rightarrow \psi)\lor 0 \rightarrow (\psi \rightarrow \phi) \lor 0} &= \interp{\psi \rightarrow \phi}\\
  \interp{(\psi \rightarrow \phi) \lor 0} - \interp{(\phi \rightarrow \psi)\lor 0} &= F(\phi) - F(\psi)\\
  (F(\phi) - F(\psi))^+ - (F(\phi) - F(\psi))^- & =F(\phi) - F(\psi)
\end{align*}
%% where $x^+ = x\lor 0$ and $x^- = (-x)\lor 0$ are the positive and
%% negative parts of $x$ respectively.
The identity $x = x^+ - x^-$ is well known for abelian lattice ordered
groups, and can be shown by $x + x^- = x + (-x)\lor 0 = (x - x)\lor(x
+ 0) = 0\lor x = x^+$.
\end{proof}

\section{Completeness}

We show completeness by relating RL to BAL \cite{Galli:04}. The
semantics of BAL is also abelian lattice ordered groups, but a
statement is interpreted as stating equality with zero. BAL has the
primitive binary operation $\rightarrow$ and unary operation
${}^+$. The former is interpreted as in RL and the latter has the
interpretation $\interp{x^+} = \interp{x}^+$, i.e.~it maps elements to
their positive parts. Thus the statement $\phi \rightarrow \psi$ in
BAL is interpreted as an assertion that $F(\psi) - F(\phi) = 0$ or
$F(\phi) = F(\psi)$. The logic has the following axioms:
\begin{align}
  \tag{BALB} &(\phi \rightarrow \psi) \rightarrow ((\chi \rightarrow \phi)
  \rightarrow (\chi \rightarrow \psi))\\
  \tag{BALC} &(\phi \rightarrow (\psi \rightarrow \chi)) \rightarrow
  (\psi \rightarrow (\phi \rightarrow \chi))\\
  \tag{BALN} &((\phi \rightarrow \psi) \rightarrow \psi) \rightarrow
  \phi\\
  \tag{BALP} &\phi^{++} \rightarrow\phi^+\\
  \tag{BALO} &((\psi\rightarrow\phi)^+
  \rightarrow(\phi\rightarrow\psi)^+)\rightarrow
  (\phi\rightarrow\psi)
\end{align}
and the following inference rules:\\
\begin{minipage}{0.49\columnwidth}
\begin{gather}
  \tag{BALMP} \frac{\phi, \phi \rightarrow \psi}{\psi},\\
  \tag{BALPI} \frac{\phi}{\phi^+},
\end{gather}
\end{minipage}
\begin{minipage}{0.49\columnwidth}
\begin{gather}
  \tag{BALG} \frac{\phi, \psi}{\phi \rightarrow \psi},\\
  \tag{BALMI} \frac{(\phi\rightarrow\psi)^+}{(\phi^+\rightarrow\psi^+)^+}.
\end{gather}
\end{minipage}
\vspace{0.3cm}\\
Note that BAL has the same expressive power as RL: a statement $x$ in
the Logic of Equilibrium is equivalent to two statements, $x$ and
$x\rightarrow 0$ in Riesz Logic. Conversely, the statement $x$ in
Riesz Logic is equivalent to the statement $(x\rightarrow 0)^+$ in the
Logic of Equilibrium: this is asserting that the negative part of $x$
is zero, which is the same as asserting that $x$ itself is positive.

Another consequence of the difference in interpretation between the
two logics is that it is not enough to show that every tautology in
BAL is a tautology in RL; we also expect their converses to hold. Our
proof of completeness is thus in three parts:
\begin{itemize}
  \item We show that the inference rules of BAL are valid in RL;
  \item We show that the axioms of BAL and their converses are
    tautologies of RL;
  \item We show that for every tautology in BAL of the form
    $(x\rightarrow 0)^+$, there is a tautology $x$ in RL.
\end{itemize}

Proofs were constructed with the help of Prover9 \cite{McCune:05}.

\subsection{Inference Rules}

Premises in BAL inference rules are stronger statements than in RL
since they are interpreted as asserting equality. Similarly, we need
to deduce two conclusions in RL for each conclusion in a BAL inference
rule in order to assert equality in RL. Specifically, given a BAL
inference rule
$$\frac{\phi_1,\phi_2,\ldots}{\psi},$$
we need the following inference rules in RL:
$$\frac{\phi_1,\phi_2,\ldots, \phi_1\rightarrow 0, \phi_2\rightarrow
  0, \ldots}{\psi}$$
and
$$\frac{\phi_1,\phi_2,\ldots,
  \phi_1\rightarrow 0, \phi_2\rightarrow 0, \ldots}{\psi \rightarrow
  0}.$$
For each rule BALR in BAL, we will refer to these two versions as
BALR+ and BALR$-$ respectively.

\begin{proof}[BALMP]
BALMP+ follows trivially from the assumption of the rule MP in RL. To see BALMP$-$:
\begin{align}
\tag{1} & \alpha \rightarrow 0 & \text{(assumption)}\\
\tag{2} & (\alpha \rightarrow \beta) \rightarrow 0 & \text{(assumption)}\\
\tag{3} & (((\phi \rightarrow \psi) \rightarrow (\chi \rightarrow \psi)) \rightarrow \omega) \rightarrow ((\chi \rightarrow \phi) \rightarrow \omega) & \text{(MP, R1a, R1a)}\\
\tag{4} & \phi \rightarrow ((\phi \rightarrow \psi) \rightarrow \psi) & \text{(MP, R1b, R1b)}\\
\tag{5} & (0 \rightarrow \phi) \rightarrow ((\alpha \rightarrow \beta) \rightarrow \phi) & \text{(MP, 2, R1a)}\\
\tag{6} & ((\alpha \rightarrow 0) \rightarrow \phi) \rightarrow \phi & \text{(MP, 1, 4)}\\
\tag{7} & (\phi \rightarrow \alpha) \rightarrow (\phi \rightarrow 0) & \text{(MP, 6, 3)}\\
\tag{8} & (\alpha \rightarrow \beta) \rightarrow (\phi \rightarrow \phi) & \text{(MP, R5a, 5)}\\
\tag{9} & \beta \rightarrow \alpha & \text{(MP, 8, R1b)}\\
\tag{10} & \beta \rightarrow 0 & \text{(MP, 9, 7)}

\end{align}
\end{proof}

\begin{proof}[BALPI]
BALPI+ follows from R2. To see BALPI$-$:
\begin{align}
& (1) && \alpha \rightarrow 0 & \text{(assumption)}\\
& (2) && \phi \rightarrow ((\phi \rightarrow \psi) \rightarrow \psi) & \text{(MP, R1b, R1b)}\\
& (3) && \phi \lor \psi \rightarrow (\phi \lor \chi) \lor \psi & \text{(RI, R2)}\\
& (4) && (\phi \lor \psi) \lor \chi \rightarrow (\psi \lor \phi) \lor \chi & \text{(RI, R3)}\\
& (5) && (\phi \lor \psi \rightarrow \chi) \rightarrow (\psi \lor \phi \rightarrow \chi) & \text{(MP, R3, R1a)}\\
& (6) && (\phi \lor \psi \rightarrow \chi) \rightarrow ((\phi \lor \psi) \lor \psi \rightarrow \chi) & \text{(MP, R4, R1a)}\\
& (7) && 0 & \text{(MP, R3, R5b)}\\
& (8) && ((\phi \rightarrow \psi) \rightarrow \chi) \rightarrow (((\psi \rightarrow \phi) \lor 0 \rightarrow (\phi \rightarrow \psi) \lor 0) \rightarrow \chi) & \text{(MP, R6a, R1a)}\\
& (9) && \alpha \lor \phi \rightarrow 0 \lor \phi & \text{(RI, 1)}\\
& (10) && (0 \rightarrow \phi) \rightarrow \phi & \text{(MP, 7, 2)}\\
& (11) && (0 \rightarrow \phi) \lor \psi \rightarrow \phi \lor \psi & \text{(RI, 10)}\\
& (12) && (0 \lor \phi \rightarrow \psi) \rightarrow (\alpha \lor \phi \rightarrow \psi) & \text{(MP, 9, R1a)}\\
& (13) && ((\phi \lor \psi) \lor \chi \rightarrow \omega) \rightarrow (\phi \lor \chi \rightarrow \omega) & \text{(MP, 3, R1a)}\\
& (14) && ((\phi \lor \psi) \lor \chi \rightarrow \omega) \rightarrow ((\psi \lor \phi) \lor \chi \rightarrow \omega) & \text{(MP, 4, R1a)}\\
& (15) && ((\phi \rightarrow \psi \lor \chi) \lor 0 \rightarrow (\psi \lor \chi \rightarrow \phi) \lor 0) \rightarrow (\chi \lor \psi \rightarrow \phi) & \text{(MP, 5, 8)}\\
& (16) && (\phi \lor \psi \rightarrow \chi) \rightarrow ((0 \rightarrow \phi) \lor \psi \rightarrow \chi) & \text{(MP, 11, R1a)}\\
& (17) && (\phi \lor \psi) \lor \phi \rightarrow \psi \lor \phi & \text{(MP, R4, 14)}\\
& (18) && \phi \lor \phi \rightarrow \psi \lor \phi & \text{(MP, 17, 13)}\\
& (19) && \alpha \lor 0 \rightarrow \phi \lor 0 & \text{(MP, 18, 12)}\\
& (20) && (\alpha \lor 0) \lor 0 \rightarrow \phi \lor 0 & \text{(MP, 19, 6)}\\
& (21) && (0 \lor \alpha) \lor 0 \rightarrow \phi \lor 0 & \text{(MP, 20, 14)}\\
& (22) && (0 \rightarrow 0 \lor \alpha) \lor 0 \rightarrow \phi \lor 0 & \text{(MP, 21, 16)}\\
& (23) && \alpha \lor 0 \rightarrow 0 & \text{(MP, 22, 15)}\\

\end{align}
\end{proof}

\begin{proof}[BALG+]
\begin{flalign*}
& (1) && \alpha \rightarrow 0 & \text{(assumption)}\\
& (2) && \beta & \text{(assumption)}\\
& (3) && \phi \rightarrow ((\phi \rightarrow \psi) \rightarrow \psi) & \text{(MP, R1b, R1b)}\\
& (4) && ((\phi \rightarrow \phi) \rightarrow \psi) \rightarrow (0 \rightarrow \psi) & \text{(MP, R5a, R1a)}\\
& (5) && (0 \rightarrow \phi) \rightarrow (\alpha \rightarrow \phi) & \text{(MP, 1, R1a)}\\
& (6) && (\beta \rightarrow \phi) \rightarrow \phi & \text{(MP, 2, 3)}\\
& (7) && 0 \rightarrow \beta & \text{(MP, 6, 4)}\\
& (8) && \alpha \rightarrow \beta & \text{(MP, 7, 5)}

\end{flalign*}
\end{proof}

\begin{proof}{BALG$-$}
\begin{flalign*}
& (1) && \alpha & \text{(assumption)}\\
& (2) && \beta \rightarrow 0 & \text{(assumption)}\\
& (3) && (((\phi \rightarrow \psi) \rightarrow (\chi \rightarrow \psi)) \rightarrow \omega) \rightarrow ((\chi \rightarrow \phi) \rightarrow \omega) & \text{(MP, R1a, R1a)}\\
& (4) && \phi \rightarrow ((\phi \rightarrow \psi) \rightarrow \psi) & \text{(MP, R1b, R1b)}\\
& (5) && ((\beta \rightarrow 0) \rightarrow \phi) \rightarrow \phi & \text{(MP, 2, 4)}\\
& (6) && (\alpha \rightarrow \phi) \rightarrow \phi & \text{(MP, 1, 4)}\\
& (7) && (\phi \rightarrow \beta) \rightarrow (\phi \rightarrow 0) & \text{(MP, 5, 3)}\\
& (8) && (\alpha \rightarrow \beta) \rightarrow 0 & \text{(MP, 6, 7)}

\end{flalign*}
\end{proof}

\begin{proof}[BALMI]
BALMI+ follows from R2. The proof of BALMI$-$ is in three parts. The
antecedent of BALMI$-$ translates to the first assumption of the
following proof, which demonstrates that if the positive part of
$\alpha \rightarrow \beta$ is less than zero, then $\beta \rightarrow
\alpha$:
\begin{flalign*}
& (1) && (\alpha \rightarrow \beta) \lor 0 \rightarrow 0 & \text{(assumption)}\\
& (2) && (\phi \lor \psi \rightarrow \chi) \rightarrow (\phi \rightarrow \chi) & \text{(MP, R2, R1a)}\\
& (3) && (\alpha \rightarrow \beta) \rightarrow 0 & \text{(MP, 1, 2)}\\
& (4) && (0 \rightarrow \phi) \rightarrow ((\alpha \rightarrow \beta) \rightarrow \phi) & \text{(MP, 3, R1a)}\\
& (5) && (\alpha \rightarrow \beta) \rightarrow (\phi \rightarrow \phi) & \text{(MP, R5a, 4)}\\
& (6) && \beta \rightarrow \alpha & \text{(MP, 5, R1b)}

\end{flalign*}
Given this, we then show that $(\alpha \lor 0 \rightarrow \beta \lor
0) \rightarrow 0$:
\begin{flalign*}
& (1) && \beta \rightarrow \alpha & \text{(assumption)}\\
& (2) && (((\phi \rightarrow \psi) \rightarrow (\chi \rightarrow \psi)) \rightarrow \omega) \rightarrow ((\chi \rightarrow \phi) \rightarrow \omega) & \text{(MP, R1a, R1a)}\\
& (3) && \phi \rightarrow ((\phi \rightarrow \psi) \rightarrow \psi) & \text{(MP, R1b, R1b)}\\
& (4) && \beta \lor \phi \rightarrow \alpha \lor \phi & \text{(RI, 1)}\\
& (5) && (((\phi \rightarrow \phi) \rightarrow 0) \rightarrow \psi) \rightarrow \psi & \text{(MP, R5b, 3)}\\
& (6) && (\alpha \lor \phi \rightarrow \psi) \rightarrow (\beta \lor \phi \rightarrow \psi) & \text{(MP, 4, R1a)}\\
& (7) && (\phi \rightarrow (\psi \rightarrow \psi)) \rightarrow (\phi \rightarrow 0) & \text{(MP, 5, 2)}\\
& (8) && (\alpha \lor \phi \rightarrow \beta \lor \phi) \rightarrow 0 & \text{(MP, 6, 7)}

\end{flalign*}
Finally, we make use of BALPI$-$ to show that we can take the
disjunction with 0 on the left-hand side.  Thus given $(\alpha
\rightarrow \beta) \lor 0 \rightarrow 0$, we can show that
$(\alpha\lor 0\rightarrow\beta\lor 0)\lor 0 \rightarrow 0$.
\end{proof}

\subsection{Axioms}

In this section, we show that the axioms of BAL are tautologies of
RL. As before, we will refer to the axiom BALA of BAL as BALA+, and
its converse as BALA$-$. Note that BALC is its own converse.

\begin{proof}[BALB+]
\begin{flalign*}
& (1) && (((\phi \rightarrow \psi) \rightarrow (\chi \rightarrow \psi)) \rightarrow \omega) \rightarrow ((\chi \rightarrow \phi) \rightarrow \omega) & \text{(MP, R1a, R1a)}\\
& (2) && \phi \rightarrow ((\phi \rightarrow \psi) \rightarrow \psi) & \text{(MP, R1b, R1b)}\\
& (3) && (((\phi \rightarrow \psi) \rightarrow \psi) \rightarrow \chi) \rightarrow (\phi \rightarrow \chi) & \text{(MP, 2, R1a)}\\
& (4) && (\phi \rightarrow \psi) \rightarrow ((\chi \rightarrow \phi) \rightarrow (\chi \rightarrow \psi)) & \text{(MP, 1, 3)}

\end{flalign*}
\end{proof}

\begin{proof}[BALB$-$]
\begin{flalign*}
& (1) && \phi \rightarrow ((\phi \rightarrow \psi) \rightarrow \psi) & \text{(MP, R1b, R1b)}\\
& (2) && ((\phi \rightarrow \psi) \rightarrow \chi) \rightarrow (((\psi \rightarrow \omega) \rightarrow (\phi \rightarrow \omega)) \rightarrow \chi) & \text{(MP, R1b, R1a)}\\
& (3) && ((\phi \rightarrow \psi) \rightarrow (\chi \rightarrow \psi)) \rightarrow (((\chi \rightarrow \phi) \rightarrow \omega) \rightarrow \omega) & \text{(MP, 1, 2)}\\
& (4) && ((\phi \rightarrow \psi) \rightarrow (\phi \rightarrow \chi)) \rightarrow (\psi \rightarrow \chi) & \text{(MP, 3, R1b)}

\end{flalign*}
\end{proof}

\begin{proof}[BALC]
\begin{flalign*}
& (1) && (((\phi \rightarrow \psi) \rightarrow (\chi \rightarrow \psi)) \rightarrow \omega) \rightarrow ((\chi \rightarrow \phi) \rightarrow \omega) & \text{(MP, R1a, R1a)}\\
& (2) && \phi \rightarrow ((\phi \rightarrow \psi) \rightarrow \psi) & \text{(MP, R1b, R1b)}\\
& (3) && (\phi \rightarrow (\psi \rightarrow \chi)) \rightarrow ((\omega \rightarrow \psi) \rightarrow (\phi \rightarrow (\omega \rightarrow \chi))) & \text{(MP, 1, 1)}\\
& (4) && (\phi \rightarrow (\psi \rightarrow \chi)) \rightarrow (\psi \rightarrow (\phi \rightarrow \chi)) & \text{(MP, 2, 3)}

\end{flalign*}
\end{proof}

\begin{proof}[BALN]
BALN+ follows from Modus Ponens on R1a and R1b; BALN$-$ follows from
Modus Ponens applied to R1b twice.
\end{proof}

\begin{proof}[BALP]
BALP+ follows from R2; BALP$-$ follows from R4.
\end{proof}

BALO+ and BALO$-$ are the only axioms we adopted unchanged in RL, as
RL6a and RL6b respectively.

\subsection{Equivalence of BAL and RL}

For every tautology of the form $(\phi \rightarrow 0)^+$ in BAL,
there is a tautology $\phi$ in RL.

\begin{proof}[Asserting Positivity]
\begin{flalign*}
& (1) && (\alpha \rightarrow 0) \lor 0 \rightarrow 0 & \text{(assumption)}\\
& (2) && ((\phi \rightarrow \psi) \rightarrow \psi) \rightarrow \phi & \text{(MP, R1a, R1b)}\\
& (3) && (\phi \lor \psi \rightarrow \chi) \rightarrow (\phi \rightarrow \chi) & \text{(MP, R2, R1a)}\\
& (4) && (\alpha \rightarrow 0) \rightarrow 0 & \text{(MP, 1, 3)}\\
& (5) && \alpha & \text{(MP, 4, 2)}

\end{flalign*}
\end{proof}

\section{Conclusion and Future Work}

We have described a new logic whose models are generalisations of
vector lattices. Vector lattices are implicit in distributional
representations of meaning used in many natural language processing
applications, and our goal is to use the new logic to combine logical
approaches to semantics with these distributional approaches. In order
to achieve this, it is likely that several enhancements will need to
be made to the logic, for example, we would probably need a
first-order and perhaps higher-order versions to accurately represent
natural language semantics.

It would also be interesting to extend the logic so that it is
complete with respect to vector lattices; to do this, we would need
some notion of multiplication by scalars.

\section*{Acknowledgments}

This work was funded by UK EPSRC project EP/IO37458/1 ``A Unified Model
of Compositional and Distributional Compositional Semantics: Theory
and Applications''.

\bibliographystyle{IEEEtran}
\bibliography{JW2012}

\end{document}